\documentclass[final]{ws-ijmpa}
\begin{document}

\markboth{RALF BERNHARD}
{Search for rare B decays at the Tevatron}

%
\catchline{}{}{}{}{}
%

\title{Search for rare leptonic B decays at the Tevatron}

\author{\footnotesize Ralf Bernhard 
}

\address{Physics Institute of the University of Z\"urich, Winterthurerstr. 190\\
CH-8057 Z\"urich, Switzerland
}

\maketitle


\begin{abstract}
Results of a search for the Flavor-Changing Neutral Current
decay $B^0_{s,d} \rightarrow \mu^+ \mu^-$ using $p\bar{p}$ collision
data at $\sqrt{s}=1.96$~TeV collected at Fermilab Tevatron
collider by the CDF and D\O~detectors are presented.
CDF reports upper limits on ${\cal B} (B^0_{s} \rightarrow \mu^+ \mu^-) \leq 7.5
\cdot10^{-7}$ and ${\cal B}(B^0_{d} \rightarrow \mu^+ \mu^- ) \leq 1.9
\cdot10^{-7}$ at the 95\% C.L. using 171 pb$^{-1}$. 
The D\O~Collaboration used 240 pb$^{-1}$ to set an even more stringent
limit on the branching ratio for $B^0_{s} \rightarrow \mu^+ \mu^-$ of $5.0\cdot
10^{-7}$ at the 95\% C.L..


\end{abstract}

\section{Introduction}

The Tevatron Run II  at Fermilab started to deliver $p\bar{p}$ collisions at
$\sqrt{s} = 1.96$ TeV in April 2002. The
$b\bar{b}$ production cross section at this center of mass energy is
very large ($\approx 150~\mu$b) when compared to the typical $e^+ e^-$
cross sections at the $\Upsilon (4S) (\approx$ 1 nb) and $Z^0$ ($\approx$
7 nb) resonances. The $b\bar{b}$ production rate at the current record
luminosity at the Tevatron of $\approx$ 10$^{32}$ cm$^{-1}$ sec$^{-1}$ 
is 10 kHz. At hadron colliders there is also the advantage of
producing all b-flavored species from the light $B^{+}_{u}$ and $B^{0}_{d}$ mesons to the heavier $B^{0}_{s}$
and $B^{+}_{c}$ meson as well as $b$ baryons such as $\Lambda^{0}_{b}$.
On the other hand, the problem of the hadronic environment is the high
level of background due a large inelastic $p\bar{p}$ cross section of
${\cal O}$(50 mb)
and the track multiplicity to up to 50 tracks/events due to the
fragmentation of the hard intersection products plus that due to the underlying
event or pile up of multiple events.

\section{Theoretical Introduction}	

The purely leptonic decay $B^0_{d,s} \rightarrow \mu^+
\mu^-$ is a Flavor-Changing Neutral Current (FCNC) process\cite{conjugated}.
In the Standard Model (SM), this decay is forbidden at the tree level and
proceeds at a very low rate through higher order diagrams.
The SM leptonic branching fraction (${\cal B}$) were calculated including QCD
corrections\cite{sm_ref2}. The latest SM predictions\cite{sm_ref3}
is, ${\cal B}(B^0_s \rightarrow \mu^+ \mu^-)=(3.42\pm 0.54)\cdot
10^{-9}$, where the error is dominated by non-perturbative hadronic
uncertainties. The corresponding leptonic branching fraction for the
$B^0_d$ is suppressed by an additional factor of $|V_{td}/V_{ts}|^2$
leading to a SM branching ratio of $(1.00\pm0.14)\cdot 10^{-10}$.
CDF reported in Run I the experimental bound for the branching fraction
of $B^0_s$ $(B^0_d)$ of ${\cal B}(B^0_s \, (B^0_d) \rightarrow \mu^+\mu^-)<2.6\cdot 10^{-6}\, (8.6\cdot 10^{-7})$ at the 95\% C.L.~\cite{cdfI}.

The decay amplitude of $B_{d,s} \rightarrow \mu^+ \mu^-$ can be
significantly enhanced in some extensions of the SM. For instance, in
the type-II two Higgs Doublet Model (2HDM), all contributions from the
neutral Higgs sector cancel out and the branching fraction depends
only on the charged Higgs mass $M_{H^+}$ and $\tan \beta$ which
defines the ratio of the vacuum expectation values of the Higgs field. The
amplitude grows as $\tan^4 \beta$~\cite{nierste}.  In the Minimal Supersymmetric
Standard Model (MSSM) however, ${\cal B}(B^0_s \rightarrow \mu^+ \mu^-) \propto\tan^6
\beta$, leading to an enhancement of up to three orders of
magnitude~\cite{Choudhury}$^,$\cite{dedes} compared to the SM, even if MSSM with minimal
flavor violation (MFV) is considered, i.e., the CKM matrix is the only
source of flavor violation. 

\section{Event Selection}

For normalization CDF used the $b$ cross section as
measured by the experiment in Run I; therefore, the
preselection criteria were driven by the measurement of the cross section with
$p_T^B >$~6~GeV and  $|y_B| < 1.0$ for events selected by dimuon triggers.  
In addition a set of standard track, muon ($p_T > 2$~GeV/$c$, $|\eta| < 0.6$) and vertex quality cuts were
applied. After restricting the invariant mass region of the dimuon pair to
4.669 to 5.696~GeV/$c^2$, 2940 events survive in a total integrated
luminosity ${\cal L}$ of 171 pb$^{-1}$.

D\O~used data collected by dimuon tiggers with two muons of opposite charge that form a
common secondary 3D-vertex with an invariant mass 
between 4.5 and 7.0 GeV/$c^2$. Each muon candidate had to have $p_T >
2.5~$GeV/$c$, $|\eta| < 2.0$ and a sufficient number of hits in the central
tracking station. To ensure a similar $p_T$ dependence of the $\mu^+\mu^-$ system in the signal and in
the normalization channel, $p_T^B$ had to be greater than 5~GeV/$c$.
After the preselection, 38,167 events survive in a integrated
luminosity of 240 pb$^{-1}$.  
 
\subsection{Discriminating variables}

Both experiments have chosen a set of similar discriminating variables
to best exploit the properties of the decay. 
The long lifetime of the $B_{s}^{0}$ meson has been used to cut against
random combinatoric background due to short lived particles. While CDF
uses a minimum proper lifetime $c\tau$ of the $B_s^0$
candidate, D\O~ is using decay length significance $L_{XY}/\delta L_{XY}$.  
The fragmentation characteristics of the $b$ quark are such that most of
its momentum is carried by the $b$ hadron. The number of extra tracks
near the B candidate therefore tends to be small. The second
discriminant was therefore an isolation variable, ${\cal I}$, of the muon pair, defined as:
\begin{equation}
   {\cal I}  = \frac{|\vec{p}(\mu^+\mu^-)|}{|\vec{p}(\mu^+\mu^-)|+ \sum\limits_{{\rm track}\,i \neq B}{ p_i(\Delta {\cal R} < 1)} }.\nonumber
\end{equation}
Here, $\sum\limits_{{\rm track}\,i \neq B}{ p_i}$ is the scalar sum over
all tracks excluding the muon pair within a cone of $\Delta {\cal R} < 1$
around the momentum vector $\vec{p}(\mu^+ \mu^-)$ of the muon pair where
$\Delta {\cal R} = \sqrt{(\Delta\phi)^2 + (\Delta\eta)^2 }$.
The CDF definition differs by using the transverse momentum of the
tracks instead of the momentum.

As a third variable the angle ($\Delta \Phi$) between the momentum vector of the muon
pair and the vector from the primary to the secondary vertex has been
used. This requirement ensures consistency between the direction of
the decay vertex and the momentum vector of the $B^0_{s}$ candidate.

\section{Limit setting procedure and optimization}

{\bf CDF}\\
For a given number of observed events, $n$, consistent with the
background estimate, $n_{bg}$, the upper limit on the branching ratio
is determined using:
\begin{equation}
 {\cal B}(B_s \rightarrow \mu^+ \mu^-) \leq
 \frac{
      N(n,n_{bg})}{2 \cdot \sigma_{B_s} \cdot \alpha
      \cdot \varepsilon_{total} \cdot \int {\cal L} dt }
\end{equation}
where N($n,n_{bg}$) is the number of candidate decays at 90\% C.L.,
estimated using a Bayesian approach\cite{CDF-limit} and incorporating
the uncertainties into the limit. 
The integrated
luminosity is ${\cal L}$, 
The $B_s^0$ production cross section is $\sigma_{B_s^0} (=
\frac{f_s}{f_u}\sigma_{B^+}$\cite{bcross} with
$f_s = {0.100}, f_u ={0.391}$\cite{pdg}) 
and $\alpha \cdot \varepsilon_{total}$ is the total
acceptance times efficiency, obtained from data and Monte Carlo. The
factor of two is necessary, since the analysis is sensitive to the
charge conjugate $b$ hadron.

The optimization was done with approximately 100 combinations of the
discriminating variables and maximized the {\sl a priori} expected limit 
which is given by the sum over all possible
observations, $n$, weighted by the corresponding Possion probability
of the expected $n_{bg}$. The optimization of the discriminating variables (c$\tau > 200 \mu
m$, $\Delta \Phi < 0.10$ rad, ${\cal I} >0.65$) predicted 1.05$\pm$0.30
background events in 171 pb$^{-1}$ .\\

\hspace{-.5cm}{\bf D\O}\\
A random grid search and an optimization
procedure\cite{punzi} was used to find the optimal cut values of the
discriminating variables by maximizing the variable
$P=\epsilon_{\mu\mu}^{B^0_s}/(a/2+\sqrt{N_{\rm bg}})$.
Here, $\epsilon_{\mu\mu}^{B^0_s}$ is the reconstruction efficiency of the
signal events relative to the preselection (estimated using MC), and
$N_{\rm bg}$ is the expected number of background events interpolated
from the sidebands. The constant $a$ is the number of standard deviations
corresponding to the confidence level at which the signal hypothesis
is tested. This constant $a$ was set to 2.0, corresponding to about the 95\% C.L.
The result ($L_{XY}/\delta L_{XY} > 18.5$, $\Delta \Phi < 0.2$ rad,
${\cal I} >0.56$) of the optimization 
leads to a background prediction of 3.7$\pm$1.1 events in 240pb $^{-1}$.

In the absence of an apparent signal a limit on the branching fraction
 ${\cal B}(B_s)$ can then be computed by normalizing the upper limit of number of events in the $B_s$ signal
region to the number of reconstructed $B^{\pm}\rightarrow  J/\psi\,K^{\pm}$ events:
\begin{equation}
\label{eq_limit2}
{\cal B}(B_s) \leq
\frac{N_{\rm ul}}{N_{B^\pm}}\cdot\frac{\epsilon_{\mu\mu K}^{B^\pm}}{\epsilon_{\mu\mu}^{B_s}}\cdot
\frac{f_{b\rightarrow B_{s}}}{f_{b \rightarrow B_{u,d}}}
{\cal B}_1(B^{\pm})\cdot {\cal B}_2(J/\psi),
\end{equation}
where $N_{\rm ul}$ is the upper limit on the number of signal decays
estimated from the number of observed events and expected background
events including systematic uncertainties using the Feldman and Cousins ordering scheme for the MC integration\cite{conrad}.
The number of observed $B^{\pm}\rightarrow J/\psi\,K^{\pm} $ events is
$N_{B^{\pm}}= 741 \pm 31\pm 22$, $\epsilon_{\mu\mu}^{B_s}$/$\epsilon_{\mu\mu K}^{B^\pm}
= 0.247 \pm 0.009 \pm 0.017$
is the efficiency ratio of the signal and normalization channels, obtained from
Monte Carlo (MC) simulations. The fragmentation ratio of a $b$ or $\bar{b}$ quark
producing a $B^0_s$ and a $B^{\pm}$ or $B_d$ is 
$f_{b\rightarrow B_{s}}/f_{b \rightarrow B_{u,d}}=0.270\pm 0.034$~\cite{pdg} and 
${\cal B}_1={\cal B}(B^\pm \rightarrow J/\psi\,K^\pm)=(1.00\pm
0.04)\cdot 10^{-3}$ and ${\cal B}_2={\cal B}( J/\psi \rightarrow \mu
\mu)=(5.88\pm 0.1)\%$~\cite{pdg} are the measured branching ratios of
the normalization channel.

\section{Results}

\begin{center}
 \begin{figure}[hbt]
  \unitlength .5cm
  \begin{minipage}{\textwidth}
    \begin{minipage}[t]{.6\textwidth}
      \begin{picture}(6,7.2)
        \leavevmode
        \epsfig{file=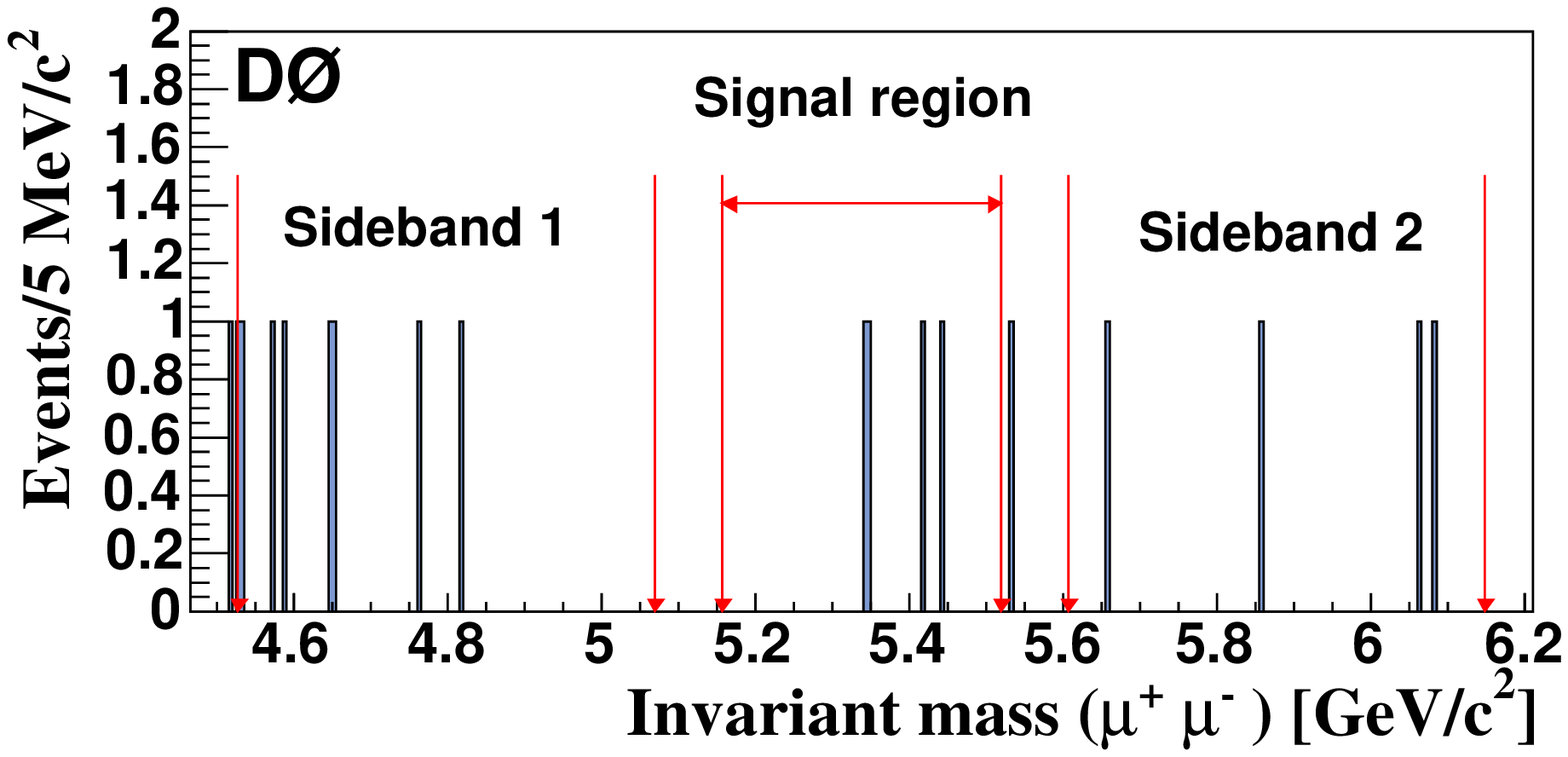,width=\textwidth}
      \end{picture}
    \end{minipage}
    \hfill
   \begin{minipage}[t]{.39\textwidth}
     \begin{picture}(6,7.2)
          \leavevmode
        \epsfig{file=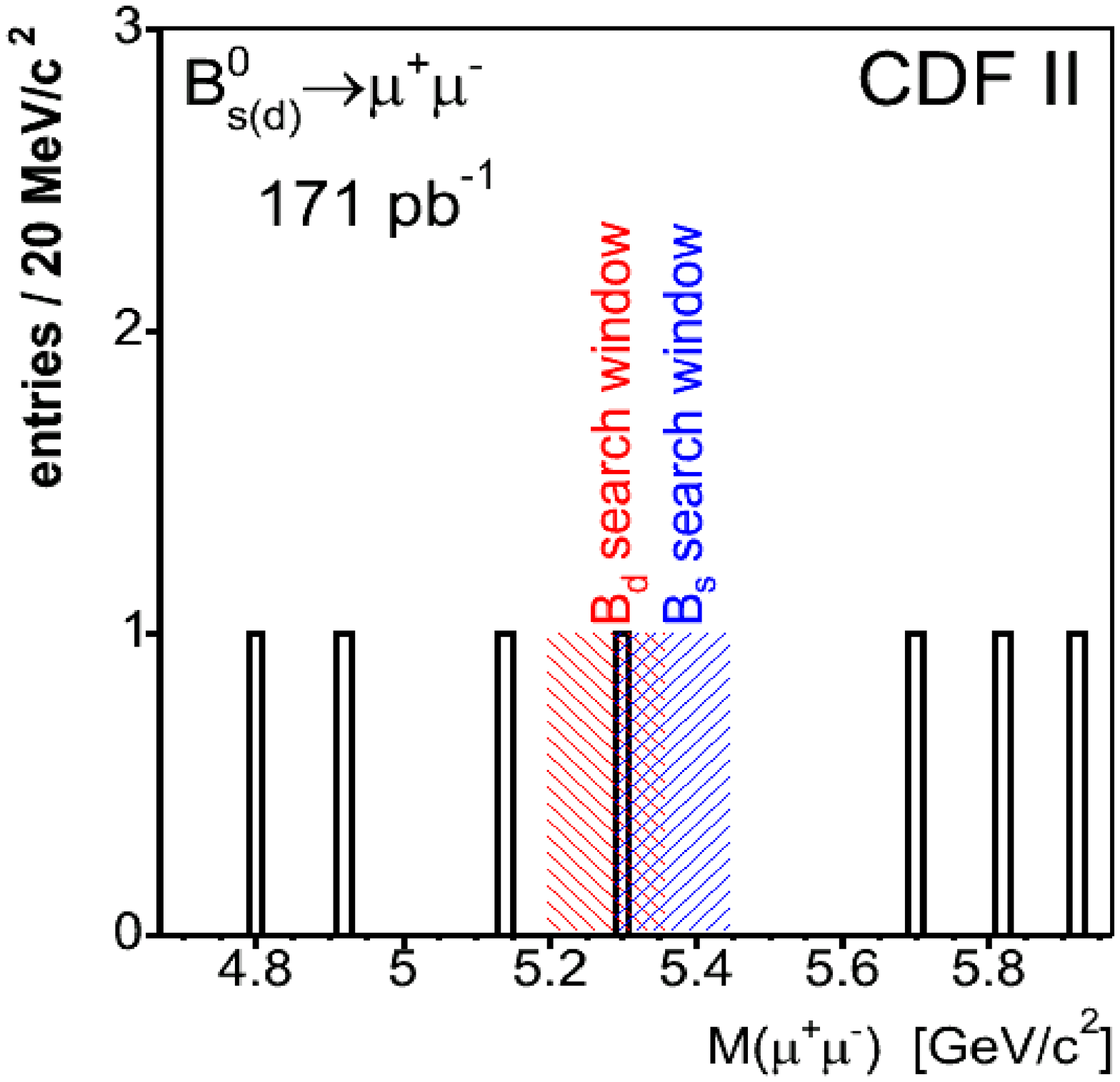,width=.9\textwidth }
     \end{picture}
      \begin{center}
      \end{center}
    \end{minipage}
  \end{minipage}
\caption{The invariant mass spectrum for the D\O~experiment (left
  side) and the CDF~experiment (right side) after
  optimized requirements on the discriminating variables.}\label{fig_mass}
\end{figure}
\end{center}

Both experiments have used a blind analysis technique, hiding the
signal region until the analysis was completed. The invariant mass
spectra after unblinding are shown in Fig.~\ref{fig_mass}. No excess
of a signal has been seen in either of the two experiments. 
CDF observed one event with an invariant mass of $M_{\mu^+ \mu^-} = 5.295$
GeV/$c^2$, thus falling into both the $B^0_s$ and $B^0_d$ mass
window. Since this is consistent with the expected background of 1.1$\pm$0.3
events in each of the mass windows ($\pm$80 MeV around the
corresponding PDG mass), 95\%(90\%) CL limits\cite{cdf_res} of ${\cal
  B}(B^0_s \rightarrow \mu^+ \mu^-) < 7.5 \cdot 10^{-7} (5.8 \cdot
10^{-7})$ and ${\cal B}(B^0_d \rightarrow \mu^+ \mu^-) < 1.9 \cdot 10^{-7} (1.5 \cdot
10^{-7})$ are found.\\
D\O~found four candidate events in the invariant mass region of $\pm$180 MeV of the expected\cite{mass}
$B^0_s$ mass, which is also consistent with the
expectation of 3.7$\pm$1.1 background events. This gives a resulting
limit\cite{d0_res} on the branching fraction at 95\% (90\%) CL of ${\cal
  B}(B^0_s \rightarrow \mu^+ \mu^-) < 5.0 \cdot 10^{-7} (4.1 \cdot
10^{-7})$. This limit assumes that there are no contributions from $B^0_d\rightarrow
\mu^{+}\mu^{-}$ decays in the search region. Any non-negligible contribution due to $B^0_d$ decays  would
make the obtained limit on the branching fraction smaller. 
The limit presented for ${\cal B}(B^0_s\rightarrow \mu^+\mu^-)$ is
therefore conservative and is even more stringent to constrain models
of new physics beyond the SM.



\begin{thebibliography}{0}

\bibitem{conjugated} Charge conjugated states are included implicitly.
\bibitem{sm_ref2}G. Buchalla and A. J. Buras, Nucl. Phys B {\bf 400}, 225 (1993); M. Misiak and
J. Urban, Phys. Lett. B {\bf 451}, 161 (1999); G. Buchalla and A. J. Buras, Nucl. Phys. B {\bf 548}, 309 (1999).
\bibitem{sm_ref3}A. J. Buras, Phys. Lett. B {\bf 566}, 115 (2003).
\bibitem{cdfI}F. Abe {\it et al.} [CDF Collaboration], Phys. Rev. {\bf D57}(1998)3811.
\bibitem{nierste}H.E. Logan and U. Nierste, Nucl. Phys. B {\bf 586}, 39 (2000).
\bibitem{Choudhury}S.~R.~Choudhury and N.~Gaur, Phys.\ Lett.\ B {\bf 451}, 86 (1999)
\bibitem{dedes}K. S. Babu and C. F. Kolda, Phys. Rev. Lett. {\bf 84}, 228 (2000); A. Dedes {\sl et al.}, FERMILAB-PUB-02-129-T.
\bibitem{nierste_prl} A. Dedes {\sl et al.}, Phys. Rev. Lett. {\bf 87}, 251804 (2001).
\bibitem{s10}T. Blazek {\sl et al.}, Phys. Lett B {\bf 589}, 39 (2004); R. Dermisek {\sl et al.}, JHEP {\bf 0304}, 37 (2003).
\bibitem{bobeth}C. Bobeth {\sl et al.}, Phys. Rev. D {\bf 66}, 074021 (2002).
\bibitem{rp_susy} R. Arnowitt {\sl et al.} Phys. Lett. B {\bf 538}, 121 (2002).
\bibitem{pdg}S.~Eidelman {\sl et al.}, Phys. Lett. B {\bf 592}, 1 (2004).
\bibitem{punzi}G.~Punzi in {\sl Proc. of the Conference on Statistical
  Problems in Particle Physics, Astrophysics and Cosmology (Phystat
  2003)}, edited by L. Lyons {\sl et al.} (SLAC, Menlo Park, CA, 2003), p. 79.
\bibitem{CDF-limit}K.~Hagiwara {\sl et al.}, Phys.~Rev.~{\bf D66}, 010001 (2002).
\bibitem{bcross}D.~Acosta {\sl et al.} [CDF Collaboration],
  Phys. Rev. {\bf D65} 052005(2002).
\bibitem{conrad}J.~Conrad {\sl et al.} Phys.~Rev.~{\bf D67}, 012002
  (2003).
\bibitem{cdf_res}D.~Acosta {\sl et al.} [CDF Collaboration],
  Phys. Rev. Lett.{\bf 93} 032001 (2004).
\bibitem{mass} To compensate for a shift in the momentum scale of the
  D\O~tracking system, the expected $B^0_s$ mass is shifted downwards
  by 30 MeV/$c^2$ with respect to the PDG mass. 
\bibitem{d0_res}Abazov, V. M. {\sl et al.} [D\O~Collaboration],
  hep-ex/0410039, submitted to PRL.
\end{thebibliography}
\end{document}